\documentclass[twocolumn,pra,amsfonts,amsmath,floatfix]{revtex4-2}
\usepackage[utf8]{inputenc}
\usepackage{enumitem}
\usepackage[colorlinks, linkcolor=blue, citecolor=blue]{hyperref}
\usepackage{graphicx}
\usepackage{adjustbox}
\usepackage[table,xcdraw]{xcolor}
\usepackage{soul}
\usepackage{multirow}
\usepackage{braket}

\begin{document}
\title{Quantum variational optimization: The role of entanglement and problem hardness}
\author{Pablo D\'{\i}ez-Valle}
\email{pablo.diez@csic.es}
\address{Instituto de F\'{\i}sica Fundamental IFF-CSIC, Calle Serrano 113b, Madrid 28006, Spain}

\author{Diego Porras}
\address{Instituto de F\'{\i}sica Fundamental IFF-CSIC, Calle Serrano 113b, Madrid 28006, Spain}

\author{Juan Jos\'e Garc\'{\i}a-Ripoll}
\address{Instituto de F\'{\i}sica Fundamental IFF-CSIC, Calle Serrano 113b, Madrid 28006, Spain}

%\date{}

\begin{abstract}
Quantum variational optimization has been posed as an alternative to solve optimization problems faster and at a larger scale than what classical methods allow. In this paper we study systematically the role of entanglement, the structure of the variational quantum circuit, and the structure of the optimization problem, in the success and efficiency of these algorithms. For this purpose, our study focuses on the variational quantum eigensolver (VQE) algorithm, as applied to quadratic unconstrained binary optimization (QUBO) problems on random graphs with tunable density. Our numerical results indicate an advantage in adapting the distribution of entangling gates to the problem's topology, specially for problems defined on low-dimensional graphs. Furthermore, we find evidence that applying conditional value at risk type cost functions improves the optimization, increasing the probability of overlap with the optimal solutions. However, these techniques also improve the performance of \textit{Ansätze} based on product states (no entanglement), suggesting that a new classical optimization method based on these could outperform existing NISQ architectures in certain regimes. Finally, our study also reveals a correlation between the hardness of a problem and the Hamming distance between the ground- and first-excited state, an idea that can be used to engineer benchmarks and understand the performance bottlenecks of optimization methods.
\end{abstract}

\maketitle

\section{INTRODUCTION}
%
% Introduction to why quantum variational optimization is important, main references
%
% Questions that remain open or futher investigation would be needed
%
% Qualitative description of our work: how we proceed
%
% Summary of the main result

In the last years, quantum variational optimization~\cite{farhi2014quantum,mcclean2016theory} has emerged as a framework that exploits the computational power of near-term intermediate scale quantum circuits~\cite{Preskill2018}. In this framework, a parametrized quantum circuit is sequentially tuned to maximize or minimize a given cost function, which is computed as a function of measured single- or multiqubit observables. While it is clear that the variational approach has broad applications in the study of complex quantum systems\ \cite{bharti2021,cerezo2020variational}, such as in quantum chemistry, there is also ongoing research on its application to NP-complete and NP-hard combinatorial optimization problems\ \cite{Willsch_2020,Nannicini_2019,headley2020approximating,Choi_2020,Hamid_2020,utkarsh2020solving,Bengtsson_2020,PhysRevApplied.14.034009,mugel2020use,fernandezlorenzo2020hybrid,osaba2021focusing,bravyi2020hybrid,Herrman_2021}, such as those represented by quadratic unconstrained binary optimization (QUBO) formulas.

In this work, we address several open questions that are open in the context of quantum discrete optimization problems. One key question is the choice of the quantum variational \textit{Ansatz}: this is the circuit that we tune to find the optimal classical solution. This circuit should efficiently sample the Hilbert space, but it also should be as simple as possible, with the minimal number of entangling operations and quantum gates, to allow their use in NISQ architectures\ \cite{haug2021capacity,holmes2021connecting}. Here, it is crucial to understand the role of entanglement and correlations, and the relation of the entanglement structure and the problem to optimize\ \cite{Harrigan2021}. These questions also tie intimately to other aspects, such as the definition of the cost function\ \cite{Li_2020,Panagiotis}, and the hybrid quantum-classical algorithms which we use to optimize the variational state\ \cite{liu2021layer,Sukin2021,Wang_2019,Liu_2019,garciasaez2018addressing,Ilya_2018,Grimsley_2019,harwood2021improving}.

In our research we address these questions with a focus on the variational quantum eigensolver (VQE) method\ \cite{Peruzzo}. We explore the performance of the VQE for a family of QUBO problems defined on random graphs with tunable density. We explore these problems by using variational \textit{Ansätze} with an entanglement structure that either mimics the original graph, or uses a random or predefined architecture. Finally, we consider both a trivial cost function, formed by the average of the QUBO Hamiltonian, or we use the average over a fixed percentage of the lowest sampled configurations, in the spirit of
the conditional Value at risk (CVaR) method\ \cite{Panagiotis}. All these problems are analyzed with a broad variety of classical optimization strategies (gradient descent, constrained optimization by linear approximation, simultaneous perturbation stochastic approximation, etc.).

Our first result is to confirm the importance of stochastic optimizers when working with an estimation of the cost function from measurements. Moreover, we conclude that in this realistic scenario, it is best to use the CVaR cost function\ \cite{Panagiotis}, rather than the average energy. The former leads to wave functions that have a greater overlap probability with the actual ground state, while the latter produces binomial distributions that can fail unconditionally, where repetition does not help. Our study also shows that for problems with a structure---i.e., graphs with low or intermediate densities---it is computationally advantageous to imitate the structure of the problem with the entanglement structure. This type of advantage vanishes when facing dense problems, such as those that are found in quantum finance scenarios. Interestingly, we find that the depth of the variational circuits is only moderately relevant. The success probability saturates for relatively shallow circuits, while the evaluation cost increases with size. This means that, for some problems, repeating the same optimization with product or weakly entangled states, can be more efficient than using more entanglement. 
In fact, we find that a product-state \textit{Ansatz} combined with the CVaR cost function yields good results that point to new quantum-inspired classical optimization methods.   
Finally, we characterized the hardness of our problems, correlating the success probability to the Hamming distance between the ground state and the first-excited state. Intuitively, this confirms the idea that for large distances, the optimizer gets trapped in low-energy solutions that are macroscopically different from the true ground state. We expect that this will be a useful indication or technique to engineer hard problems, as well as a hint that we need to engineer quantum algorithms that enhance the probability of macroscopic tunneling between solutions.

This paper is structured as follows: In Sec. \ref{sec_formulationoftheproblem} we describe the combinatorial optimization problems that we focus on in this work, QUBO problems, and its straightforward mapping as a Hamiltonian in order to be solved by a variational hybrid method. In Sec. \ref{sec_methodology} we introduce the details of the resolution methods. Later, in Sec. \ref{sec_results} we present our numerical results obtained from ideal simulations with exact state calculation and finite sampling.  These results highlight the improvement introduced by a CVaR style cost function (Sec. \ref{subsec_cvarcqeperformance}), they show the impact of the structure of the variational form (Sec. \ref{subsec_entanglinglayereffect}), they give us intuition into how the problem characteristics can affect the performance of the algorithm (Sec. \ref{subsec_hardnessanalysis}), and they compare the efficiency of \textit{Ans\"atze} based on entangled states with that based on product states (Sec. \ref{subsec_productstatevsentangledstate}). Finally, we conclude by summarizing the main ideas of the paper in Sec. \ref{sec_conclusions}.

\section{Quadratic unconstrained binary optimization problems on random graphs}
\label{sec_formulationoftheproblem}

Combinatorial optimization is one of the most common areas studied in NISQ quantum computing and quantum annealing.  A broad spectrum of combinatorial optimization problems can be represented as QUBO problems, with a cost function $E$ that involves products of two Boolean variables:
\begin{equation}
    E(\vec{x}) = \sum_{i,j} x_i Q_{i j} x_j \;\;;\;\;\; x_i \in \{0, 1\}.
\label{QUBO}
\end{equation}
Here $\vec{x}$ is an $N$-vector of binary variables, and $Q$ is an $N$-by-$N$ square symmetric matrix of coefficients.  QUBO problems have been extensively studied in the literature \cite{QUBO1,QUBO3}; for instance, solving optimization problems on graphs, resources allocation problems, clustering problems, satisfiability, sequencing and ordering problems, facility locations problems, and various forms of assignment problems.

Thanks to the matrix structure, QUBO problems can be regarded as the representation of an undirected graph with $N$ vertices, connected by undirected edges $i\leftrightarrow{j}$ that have associated weights $Q_{ij}=Q_{ji}.$ A classical solution to the QUBO problem corresponds to labeling each vertex with $0$ or $1,$ so as to maximize or minimize the cost function $E$ [see Fig. \ref{variationalformRY}(b)]. In this work we classify QUBO problems according to the size of the corresponding graph, given by the number of vertices $N$, and the \textit{density of the graph,} defined as the ratio
\begin{equation}
  D = \frac{2E }{N(N-1)} ,
\end{equation}
between the number of edges $E$ and the maximum number of potential connections $N(N-1)/2$.

A QUBO problem can always be mapped to an Ising spin model. Thereby the optimization of the cost function $E$, Eq.\ \eqref{QUBO}, becomes a search of the minimum-energy state of an Ising Hamiltonian for an $N$-qubit system where each qubit represents one vertex of the graph.The transformation from a QUBO to an Ising formulation is implemented through the change of variables $x_i \rightarrow \frac{1}{2}(1 + \sigma^z_i )$, where $\sigma^z_i$ is the Pauli $Z$ matrix acting on the $i$-th qubit. We can express the previous function as
\begin{equation}
\begin{split}
& f(\sigma^z_i) = -\frac{1}{2} H + \frac{1}{4}\sum_{i,j} Q_{ij}, \\
& \mbox{ with } H = \sum_{i<j} J_{ij} \sigma^z_i \sigma^z_j +  \sum_{i} h_{i} \sigma^z_i ,
\end{split}
\label{Hamiltonian_IS}
\end{equation}
using the coupling matrix $J_{ij}=-Q_{ij}$ and the magnetic fields $h_{i}=-\sum_j Q_{ij}$. In this paper we only consider frustrated systems, where the matrices $J$ and $Q$ combine both positive and negative coefficients, and the optimization is potentially harder.

\section{Methodology}
\label{sec_methodology}
\subsection{Hybrid quantum-classical algorithms}
\label{sec_Hybridquantum-classicalalgorithms}
Hybrid quantum-classical algorithms are those that combine the use of quantum and classical computers to solve a computational problem. In quantum variational algorithms\ \cite{Nikolaj, mcclean2016theory}, the quantum computer is used to prepare and characterize a complex wave function, defined in terms of parametrized quantum operations
\begin{equation}
  \ket{\Psi (\vec{\theta})} = U_L(\theta_L) ... U_1(\theta_1) \ket{\Psi_0}.
  \label{ansatzintro}
\end{equation}
The real values $\theta_i$ determine how this state is to be constructed in a quantum computer by composing unitary operations $U_i(\theta_i).$ The parameters of this wave function are tuned according to some classical algorithms, until the wave function maximizes or minimizes a measure of ``fitness'' or cost function. In this work the fitness function will be a QUBO problem, and we will consider two hybrid algorithms: the original VQE and its modification based on CVaR measures.

\subsubsection{Variational quantum eigensolver algorithm.}
\label{sec_variationalquantumeigensolveralgorithm}

The variational quantum eigensolver, or VQE\ \cite{Peruzzo}, is a hybrid quantum-classical algorithm that seeks the minimal eigenvalue $\lambda_\text{min}$ and corresponding eigenstate of a problem Hamiltonian $H,$ such as the QUBO models described above\ \eqref{Hamiltonian_IS}. The algorithm relies on a parametrized family of wave functions\ \eqref{ansatzintro}. The variational principle states that the average energy of this state is strictly larger or equal to the energy of the ground state we seek:
\begin{equation}
  \lambda_{\rm min} \leq E(\vec\theta)= \braket{\Psi(\vec\theta)| H |\Psi(\vec\theta)}.
\end{equation}
Thus, if our parametrization\ \eqref{ansatzintro} is dense in the Hilbert space, we can obtain a very good approximation to the ground state by solving the problem
\begin{equation}
  \vec\theta_\text{opt} =  \text{argmin}_{\vec\theta}E(\vec\theta).
\end{equation}
VQE solves this problem by iteratively preparing the wave function $\ket{\Psi(\vec\theta^{(k)})}$ and measuring its energy $E(\vec\theta^{(k)})$ in the quantum computer. The VQE relies on a classical optimization algorithm---gradient descent, SPSA, Adam, etc---to build a progressively optimized sequence of parameters $\{\vec\theta^{(k)}\}$ that classically aims to minimize our black-box functional $E(\vec\theta).$

VQE is particularly well suited to address QUBO problems, where the energy functional\ \eqref{Hamiltonian_IS} is written as a sequence of mutually commuting observables. In this case, to solve a QUBO problem with $N$ variables, the algorithm relies on a quantum computer with $N$ qubits that are prepared in some trial state $\ket{\Psi(\vec\theta)}.$ The energy of this state is approximated by simultaneously measuring all qubits in the Pauli $z$ basis. This produces a string of $N$ values $\{s_i\in\{-1,1\}\},$ which can be translated to a bit string $\{x_i\}$ and a QUBO cost function\ \eqref{QUBO} that we label $E_n$. By averaging these values over $K$ repetitions, we obtain a stochastic approximation
\begin{equation}
  \bar{E}_K(\vec\theta) := \frac{1}{K}\sum_{k=1}^{K}E_k \to E(\vec\theta),\,K\to\infty
  \label{samplemean}
\end{equation}
that is iteratively optimized using the classical methods mentioned before.

\subsubsection{Conditional value at risk-variational quantum eigensolver}
\label{sec_cvarvqe}
A problem with the VQE is that it regards all outcomes of the measurement process with equal importance, even outlier states that may have an energy far away from our goal. If we call $P(E;\vec\theta)$ the distribution of energies associated with the states we build, the VQE cost function is
\begin{equation}
  \bar{E}_K(\vec\theta)\simeq \mathbb{E}[E(\vec{\theta})]=\braket{\Psi(\vec{\theta})| H |\Psi(\vec{\theta})} = \int_{-\infty}^{\infty} P(E;\vec{\theta})E dE.
  \label{expectedvalue}
\end{equation}
Recently, Barkoutsos et al.\ \cite{Panagiotis} proposed a variation of VQE that replaces the cost function with a new estimator that is inspired by the CVaR or the \textit{expected shortfall} definition applied in finance\ \cite{Acerbi}. This new estimator only works with a fraction $\rho\in(0,1)$ of the bit-strings that the quantum computer produces, selecting them to be those that have the lowest energy. If we sort all sampled energies $\{E_1\leq E_2\leq \ldots\leq E_K\},$ and we take the fraction $\rho$ of lowest energy outcomes, 
\begin{equation}
    \textnormal{CVaR}_\rho(E(\vec{\theta})) \approx \frac{1}{\lfloor\rho K\rfloor}\sum_{k=1}^{\lfloor\rho K\rfloor}E_k.
    \label{samplemeanCVaR}
\end{equation}
In the limit of many measurements, the CVaR-VQE estimator approximates the cost function
\begin{equation}
\begin{split}
    \textnormal{CVaR}_\rho(E(\vec{\theta})) &= \mathbb{E}[E(\vec{\theta})|E\leq F_{E}^{-1}(\rho)] = \\ & =\frac{1}{\rho} \int_{-\infty}^{E_\rho(\vec{\theta})} P(E;\vec{\theta})E dE\;\;,
\end{split}
\end{equation}
where $F_E$ represents the cumulative density function of $E$ and $E_\rho(\vec{\theta})$ is computed as
\begin{equation}
    \int_{-\infty}^{E_\rho(\vec{\theta})} P(E;\vec{\theta})dE = \rho\,.
\end{equation}
The intuition behind the idea is that improving the properties of the best measurement outcomes may be more efficient than improving the average of all outcomes, by concentrating the probability of the wave function in the lowest-energy states.

\subsection{Quantum variational wave functions}
\label{sec_ansatzs}
A crucial ingredient in all variational algorithms is the choice of parametrized wave function or \textit{Ansatz}\ \eqref{ansatzintro}. Typically, these \textit{Ans\"atze} are physically motivated and combine local gates with entangling operations that correlate the qubits in a hardware efficient manner. More generally, the choice of gates and the topology of the entangling operations can affect the efficiency of the outcome. In this study we focus on four types of variational states: three types wave functions that are created by the action of control $Z$ gates, and also a family of cheap product states where no correlation is required.

\begin{figure}
\centering
\includegraphics[width=1\linewidth]{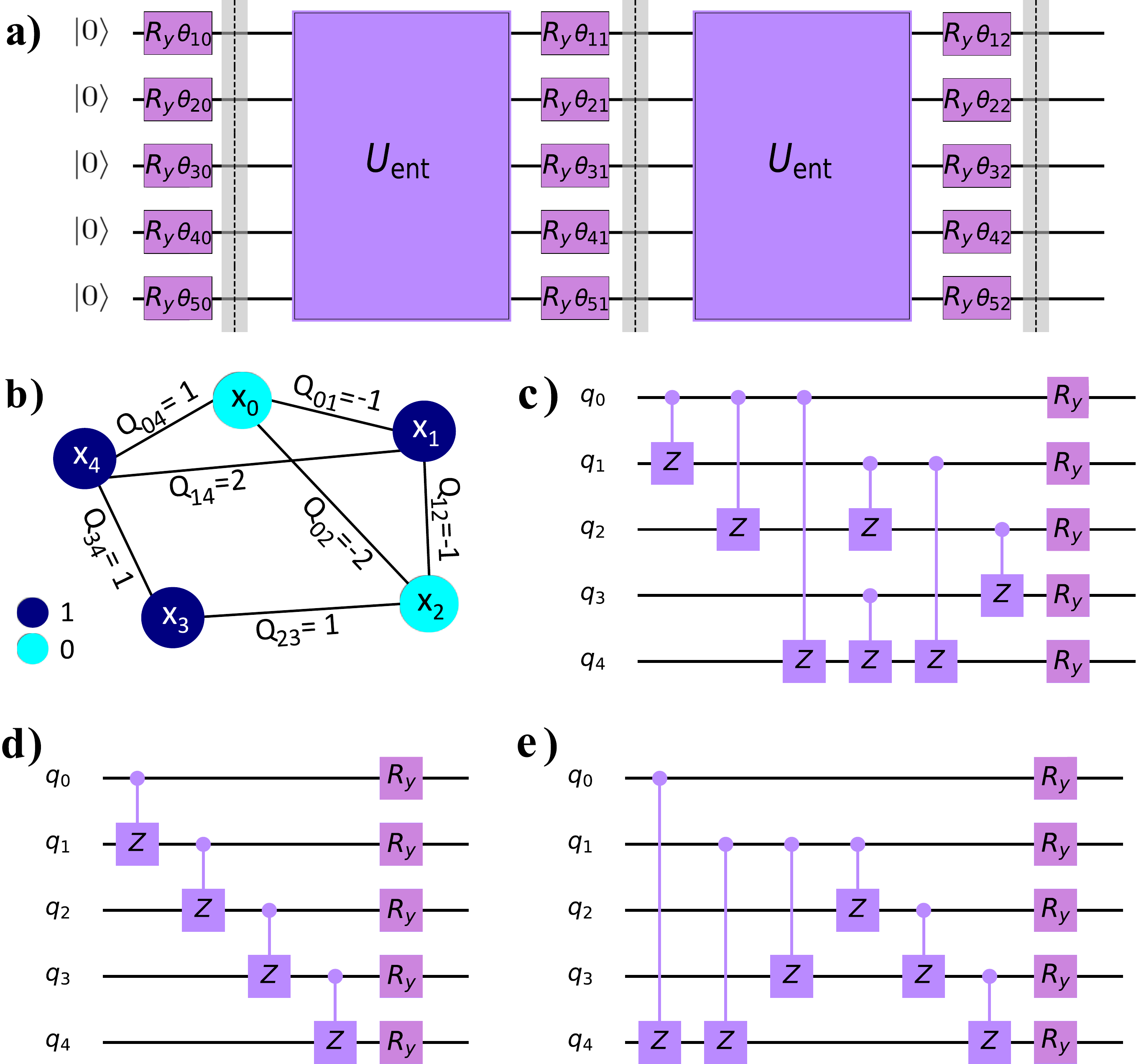}
\caption{(a) Variational form based on single-qubit $\sigma^y$ rotations for five qubits with two layers and fifteen variational parameters. (b) Example of a particular optimized graph. (c) Entangling layer compatible with the graph structure. (d) Entangling layer with a linear entanglement. (e) Entangling layer with a random entanglement.}
\label{variationalformRY}
\end{figure}

\subsubsection{Entangled states}
We start by considering a set of entangled variational wave functions as originally proposed for the VQE method\ \cite{mcclean2016theory},
\begin{equation}
\label{ansatz}
 \ket{\Psi (\vec{\theta})} = U(\vec{\theta}) \ket{0},
\end{equation}
with
$U(\vec{\theta}) =
\prod_{l=1}^{L}\left[\prod_{n=1}^{N}e^{i\theta_{n l}\sigma^y_n} U_{\textnormal{ent}}\right]\prod_{n=1}^{N}e^{i\theta_{n 0}\sigma^y_n}$.

This wave function combines $L$ layers of entangling gates $U_\text{ent}$ with real-valued rotations generated by the qubits' Pauli $y$ matrices $\sigma^y_n.$ The $N\times (L+1)$ angles of these rotations $\theta_{nl}$ are the variational parameters of the wave function (cf. Fig.\ \ref{variationalformRY}). The entangling unitary, $U_{\rm ent}$, determines the correlation structure of the wave function. We will test three such structures, built from two-qubit control $Z$ gates:
\begin{enumerate}[label=(\roman*)]
    \item Linear entanglement: The entangling gates are independent of the structure of the QUBO problem.
    Every entangling layer is made up of two-qubit control $Z$ gates between each qubit with its nearest-neighbor qubit in a linear quantum processor topology [see Fig. \ref{variationalformRY}(d)].
    \begin{equation}
        U_{\textnormal{ent}}= \prod_{i} e^{i\frac{\pi}{4}\left(\mathbb{I}-\sigma^z_i\right)\left(\mathbb{I}-\sigma^z_{i+1}\right)} \;\;,
    \end{equation}
    where $\sigma^z$ is the Pauli $Z$ matrix acting on the $i$th qubit.
    \item Compatible entanglement: Entangling gates are chosen to reflect the structure of the QUBO problem. Every entangling layer is made up of two-qubit control $Z$ gates between each qubit with nearest neighbors in the QUBO graph [see Fig. \ref{variationalformRY}(c)].
    \begin{equation}
        U_{\textnormal{ent}}= \prod_{\langle i,j \rangle} e^{i\frac{\pi}{4}\left(\mathbb{I}-\sigma^z_i\right)\left(\mathbb{I}-\sigma^z_{j}\right)} \;\;,
    \end{equation}
    where $\langle i,j \rangle$ denote that $i$ and $j$ are nearest neighbors, i.e., $Q_{i j}\neq0$ with $Q$ being the QUBO matrix.
    \item Random entanglement: Every entangling layer is made up of two-qubit control $Z$ gates between random qubits. The number of entangled qubits is given by the number of nonzero weights in the QUBO graph, i.e., it is not a fully random entanglement but it is dependent on the density of the problem [see Fig. \ref{variationalformRY}(e)].
    \begin{equation}
        U_{\textnormal{ent}}= \prod_{i,j} e^{i\frac{\pi}{4}\left(\mathbb{I}-\sigma^z_i\right)\left(\mathbb{I}-\sigma^z_{j}\right)} \;\;,
    \end{equation}
        where $i\neq j$ are $E$ random pairs.
        %with $E$ the total number of edges of the graph.
\end{enumerate}
In our variational algorithms we typically initialize the VQE to a uniform superposition of all states in the computational basis, using Hadamard-like rotations $\theta_{n 0} = \pi / 4,$ and making all remaining layers weak perturbations over this state, $\theta_{n l}\approx 10^{-2}$ for $l\geq 1$.

\subsubsection{Product state method}
In this work we also explore fully separable variational states, built from $\sigma^y$ rotations of the $N$ qubits
\begin{equation}
  \label{product_state}
  \ket{\Psi (\vec{\theta})} = \bigotimes_{j=1}^N \ket{\psi (\theta_j)},
\end{equation}
with
 $\ket{\psi (\theta_j)}  = \cos{\theta_j} \ket{0} + \sin{\theta_j} \ket{1}$.

The state \eqref{product_state} can be understood as a borderline case of the variational form \eqref{ansatz} exposed in the previous section in which the number of layers $L$ is 0. The elimination of the entangling gates makes the state hardware efficient and more accurate, but it also makes the whole variational procedure efficiently simulable in a classical way. We can regard this family of states as a new type of classical algorithms, much like the spin-vector quantum Monte Carlo methods that have been used to benchmark quantum annealers\ \cite{Smolin2014,Shin2014}.

\subsection{Efficiency indicators}
\label{subsec_efficiencyindicators}

\begin{figure}
\centering
\includegraphics[width=\linewidth]{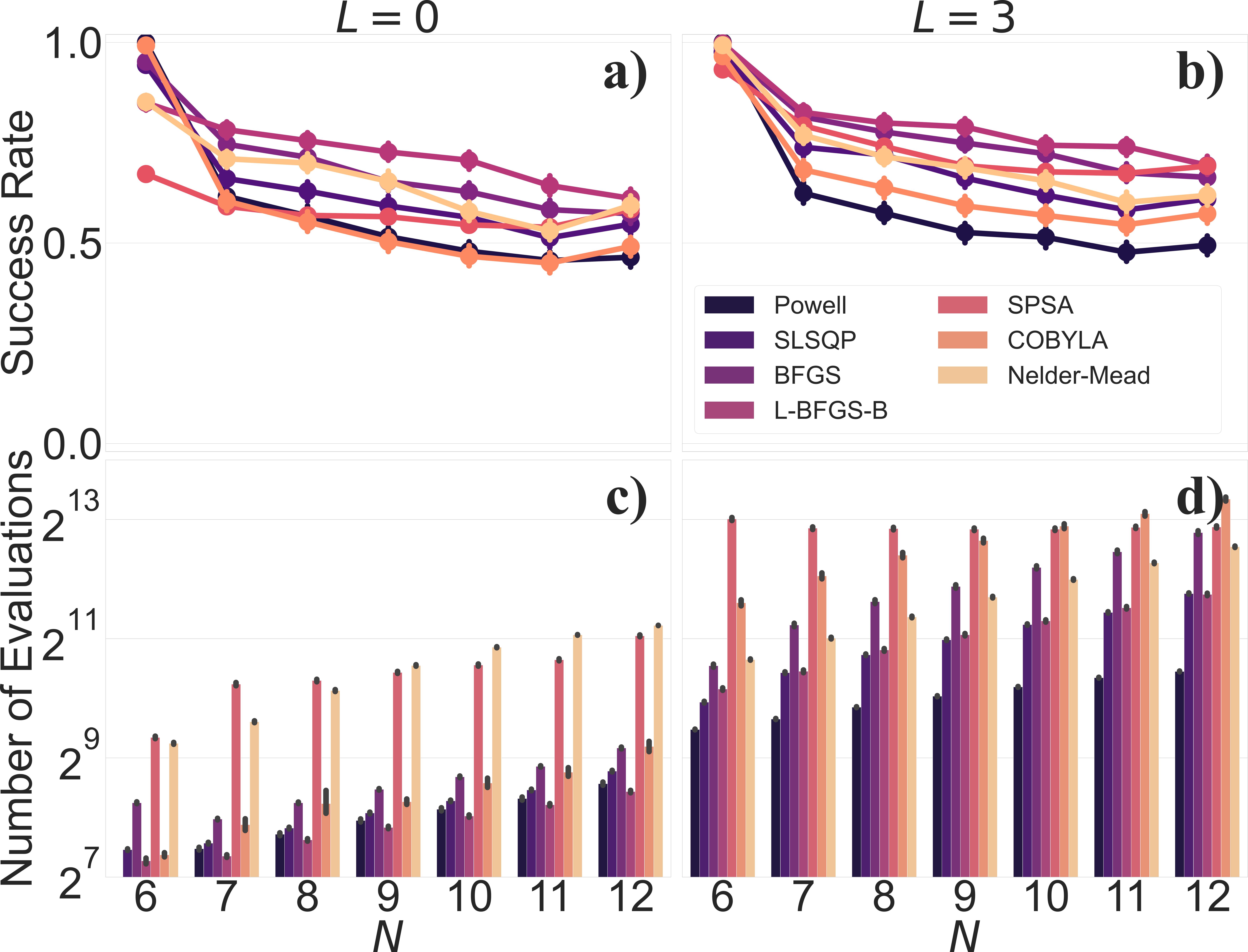}
\caption{Comparison of VQE performance with a variety of classical optimization methods to optimize the variational parameters using exact quantum states resulting from simulation. From left to right we increase the number of layers $L$ of the \textit{Ansatz}. On the $x$ axis we plot the size of the problem: (a),(b) success rate, (c),(d) objective function evaluations needed to converge. The results show the average of 1600 instances and a 95\% confidence interval.}
\label{classicaloptimizers}
\end{figure}

\begin{figure}
\centering
\includegraphics[width=\linewidth]{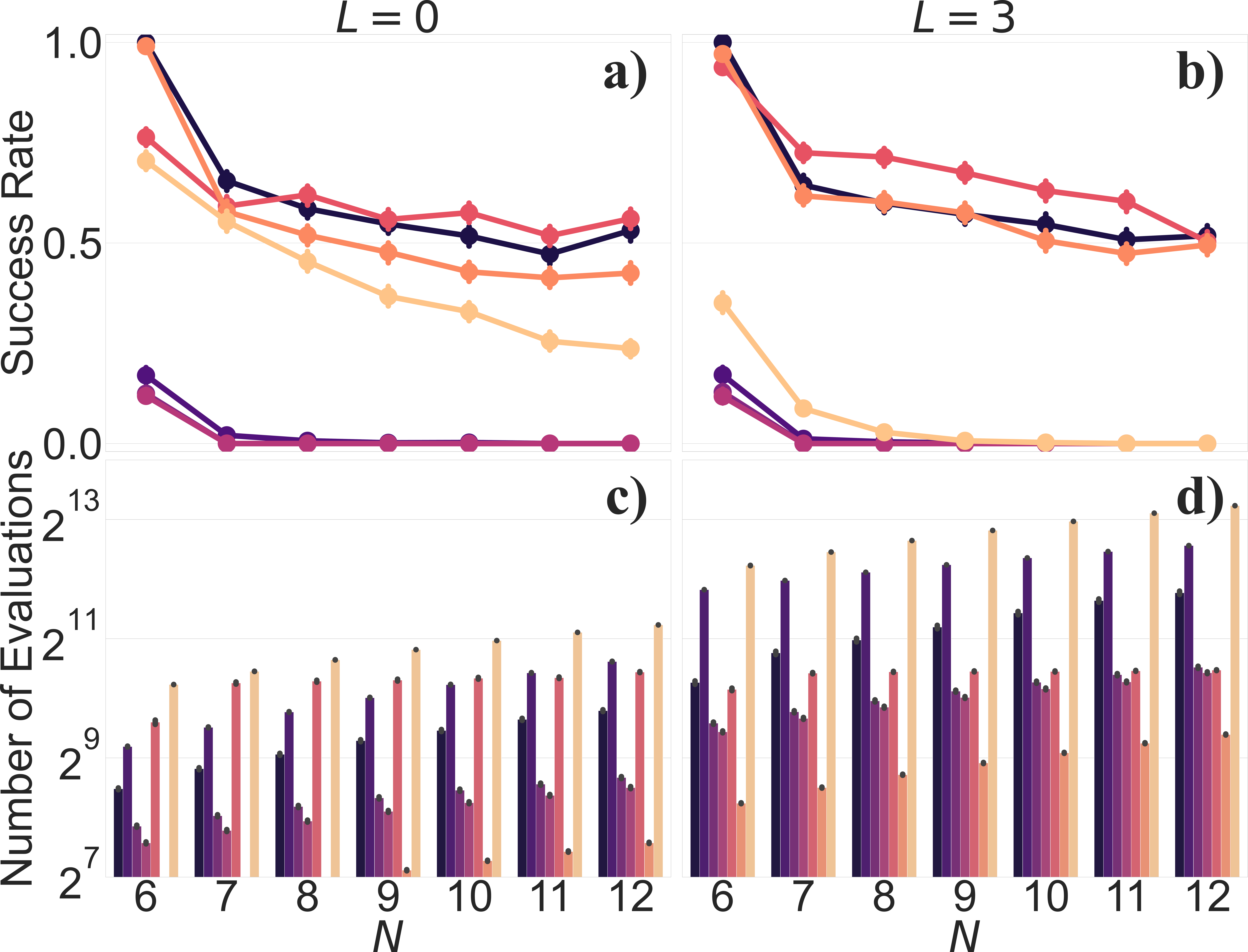}
\caption{Comparison of VQE performance with a variety of classical optimization methods to optimize the variational parameters using approximate quantum states computed by a finite set of measurements on the qubits resulting from simulation. From left to right we increase the number of layers $L$ of the \textit{Ansatz}. On the $x$ axis we plot the size of the problem. (a),(b) success rate, (c),(d) objective function evaluations needed to converge. The results show the average of 1600 instances and a 95\% confidence interval.}
\label{classicaloptimizers_samples}
\end{figure}

To study the efficiency of the algorithms we use the algorithm's runtime and the success as indicators. The algorithm's speed is quantified by the number of the cost function evaluations needed to converge to an optimal solution (global or local). Note that this is also the number of queries made to the quantum processor, the most time-consuming step in the whole process.

The algorithm's success is quantified by the probability that the final wave function $\ket{\Psi_\text{out}}$ has a significant overlap with the target state $\ket{\Psi_\text{solution}}$. 
This target state is previously computed by a classical method. 
We deem the algorithm successful $S=1$ whenever this probability exceeds a fixed cut-off $\beta$:
\begin{equation}
    S \equiv \left\{
\begin{array}{ll}
0 & \text{ if } |\langle\Psi_{\rm out} | \Psi_{\rm solution} \rangle|^2 < \beta \;\;,\\
 1 & \text{ if } |\langle\Psi_{\rm out} | \Psi_{\rm solution} \rangle|^2 \geq \beta \;\;.
\end{array}
\right.
\end{equation}
Let us point out that the algorithm could converge to a local minimum of the cost function quite different from $\ket{\Psi_{solution}}$, since the classical optimizers use standards convergence criteria based on the change of the cost function value in each optimization step. Therefore the convergence of the algorithm does not guarantee $S=1$. 
Given a set of problems, we call the \textit{success rate} the average of $S$ over all such instances. The value of $\beta$ is somewhat arbitrary, but once fixed it lower bounds the probability of obtaining the right solution by measuring over the final state $k$ times
\begin{equation}
  P(\Psi_\text{solution}|S=1) \geq 1 - (1 - \beta)^k.
\end{equation}
In our simulation we set $\beta=0.1$ so that the success of the algorithm implies we obtain the right state with $99.997\%$  probability after $k=100$ measurements of the variational state.

\subsection{Classical benchmark of the algorithms}
\label{sec_detailsofcomputationalexperiments}

Actual variational quantum algorithms are expected to run on real quantum hardware. However, the goal of this work is to perform a neutral benchmark focused on the influence of variational states and cost functions in the algorithm, leaving aside experimental imperfections, such as noise or disconnected qubits. For this reason, we conduct our benchmarks by using a classical simulator of an ideal quantum computer. All simulations as well as the optimization of the variational parameters were performed on nodes with two Intel Sandybridge E5-2670 processors, each one with eight cores operating at 2.6 GHz and a cache of 20MB and 64 GB of RAM memory (i.e., 4 GB/core) running Centos 7.4.

Another important ingredient in these simulations is whether we consider a finite number of measurements---i.e., $K$ is finite in Eqs.\ \eqref{samplemean} and \eqref{samplemeanCVaR}---, or whether we study the limit of infinite measurements, where the quantum uncertainty is eliminated. In this work we begin with the second set of results, which are obtained by simulating the complete wave function and evaluating the exact expectation values $E(\vec\theta)$ or $\text{CVaR}_\rho(E)$. These results are later on complemented by stochastic simulations that emulate a finite number of measurements and the randomness of a quantum computer.

\subsection{Choice of classical optimizer}
\label{sec_classicaloptimizers}

In our study of the variational method we have tested seven different optimizers: modified Powell's conjugate direction method \cite{Powell}, sequential least squares programming (SLSQP) \cite{SLSQP}, Broyden–Fletcher–Goldfarb–Shanno (BFGS) \cite{BFGS1,BFGS2,BFGS3,BFGS4} and its limited-memory version (L-BFGS-B) \cite{L-BFGS-B}, the simultaneous perturbation stochastic approximation (SPSA) \cite{SPSA}, the constrained optimization by linear approximation \cite{COBYLA} and the Nelder-Mead algorithm \cite{Nelder-Mead}. We use the optimization methods straightforwardly provided by Scipy 1.3.2 \cite{Scipy}, except for SPSA, which is a homemade implementation.

Before our full study of the variational algorithms, we benchmarked these optimizers in a limited set of QUBO problems. Figure\ \ref{classicaloptimizers} shows the success rates and number of evaluations of these methods vs the number of qubits. The results have been averaged over 1600 randomly generated problems, optimized in the limit of infinite number of measurements---i.e., wave function simulations---. Figure\ \ref{classicaloptimizers_samples} shows similar results, but emulating 9000 measurements of a quantum computer per evaluation.

These figures show a stark contrast in the actual performance of the optimizer. Gradient-free optimizers such as SPSA, COBYLA, Powell, and Nelder-Mead perform well even when the information of the objective function is not complete and its computation presents some stochasticity. Gradient-based optimizers such as SLSQP, BFGS, and L-BFGS-B perform very well in wave function simulations, even outperforming the other methods. However, these algorithms fail when the cost function is evaluated with some uncertainty, becoming trapped in local minima due to an imperfect estimate of the optimal descent direction. Consequently, in the remainder of this work we use L-BFGS-B when working with wave function simulations, while we use SPSA in scenarios with a finite number of measurements.

\section{Results and discussion}
\label{sec_results}

Let us now show the main results we obtained by benchmarking the VQE and CVaR-VQE algorithms on random instances of QUBO problems. These problems were generated by creating regular and unstructured random graphs with  the Python package \textit{networkx}\ \cite{networkx}, and assigning to the edges of the graph a uniformly sampled random integer weight on the $[-10,10]$ interval. We performed the experiments for various numbers of qubits $N$ and graph densities $D$, generating $N_\text{ins}=1600$ instances of each configuration.
A different set of 1600 instances was used in each experiment. 
The final results are shown in Figs. \ref{classicaloptimizers}-\ref{hardnesssuccessandeval_sampling} with 95\% confidence intervals, which are obtained by resampling the output data with the bootstrap method\ \cite{Efron1979,Efron1993book}.

\subsection{CVaR-VQE vs. standard VQE performance}
\label{subsec_cvarcqeperformance}

\begin{figure}
\centering
\includegraphics[width=1\linewidth]{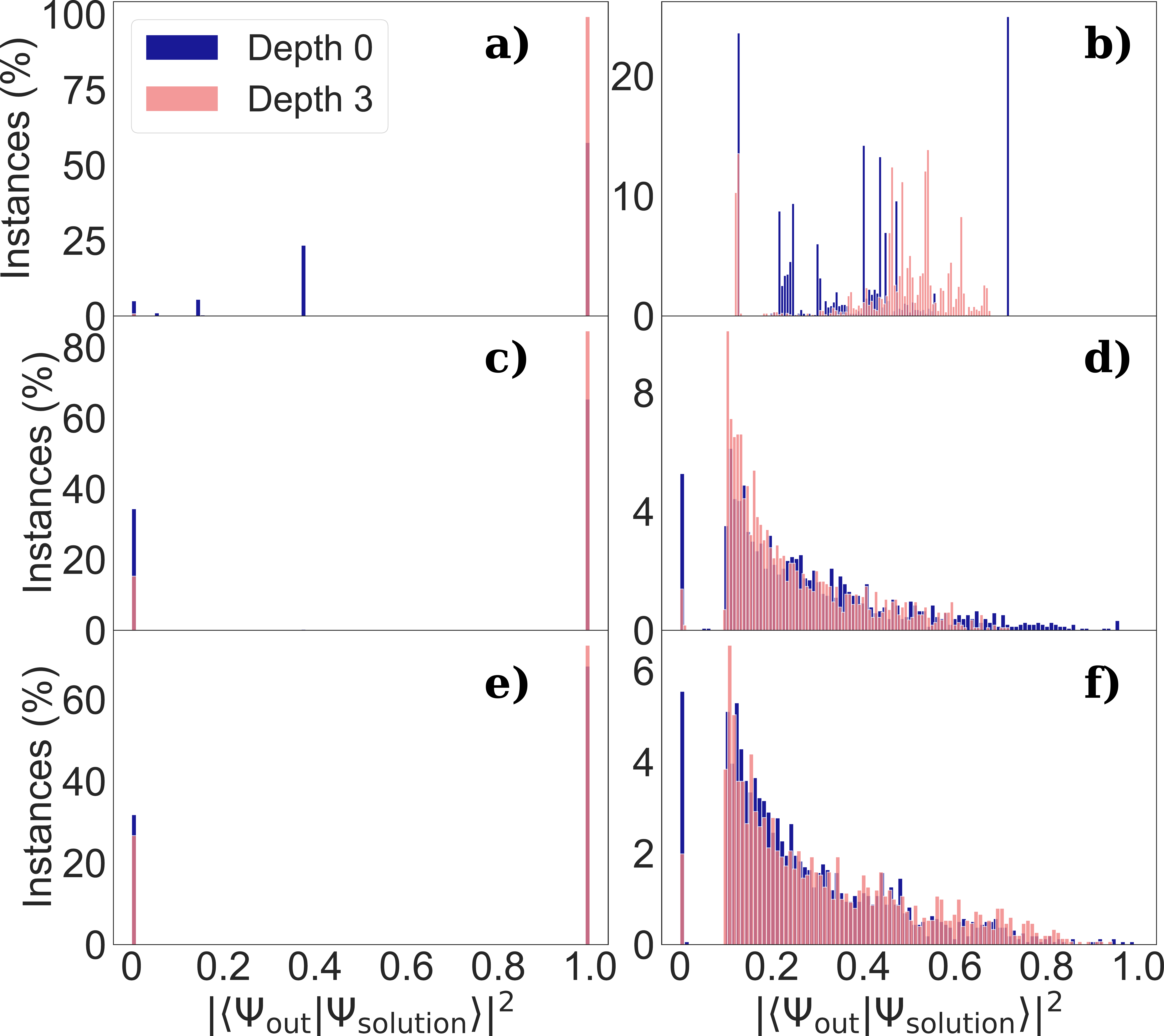}
\caption{Distribution of the overlap of the final state $\ket{\Psi_{\rm out}}$ resulting from the convergence of the algorithm with the exact ground state $| \Psi_{\rm solution} \rangle$ computed classically. We plot the percentages obtained from $N_{\rm ins} = 1600$ instances for an \textit{Ansatz} with $L=0$ and $L=3$ entangling layers, $N=12$ qubits, and using L-BFGS-B as classical optimizer. We show on the left the results from VQE standard algorithm, and on the right the results from CVaR-VQE with $\rho = 10\%.$ From top to bottom, the density is (a), (b) $0.045$; (c), (d) $0.258$; and (e), (f) $0.894.$ In this paper we consider successful trials those whose final overlap is at least 0.1.}
\label{distributioncvarvqe}
\end{figure}

\begin{figure}
\centering
\includegraphics[width=1\linewidth]{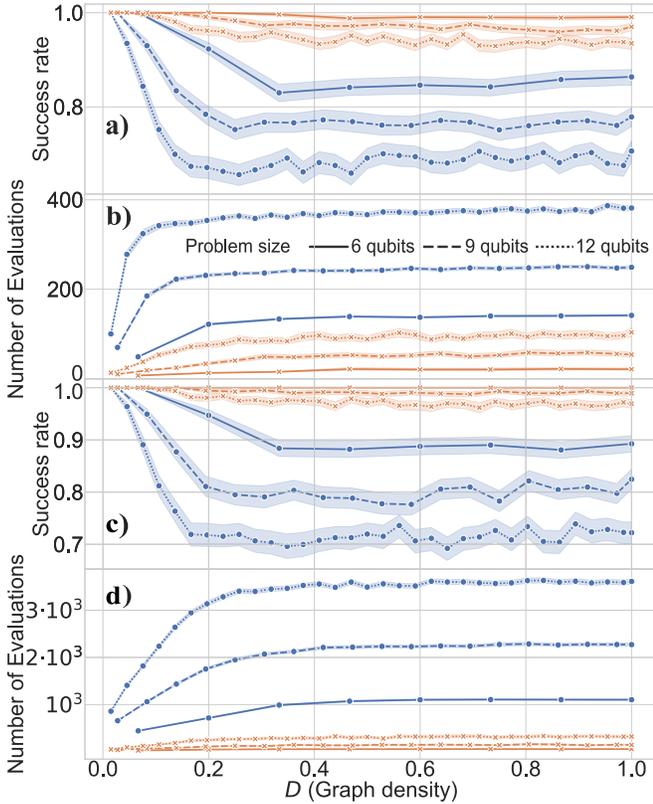}
\caption{Idealized performance of standard VQE (blue circles) and CVaR-VQE with $\rho = 10\%$ (orange X symbols), as a function of graph density $D$ for a classical simulation with the full wave function $(N_\text{shots}\to\infty)$. We compare (a), (b) a product state with (c), (d) an entangled state with $L=3$ layers for problems with different $N=6$, 9 and $12$ qubits. We plot (a), (c) the success rate and (b), (d) the number of function evaluations vs the density of the underlying graph. The lines show the average over $N_{\rm ins}=1600$ problem instances, surrounded by a colored 95\% confidence interval.}
\label{cvarvsvqe}
\end{figure}

First, we present a comparison between the performance of standard VQE and CVaR-VQE (with $\rho =10\%$), using an \textit{Ansatz} with three entangling layers
($L = 3$) and a product state wave function.

\textbf{Final state wave function: CVaR-VQE enhances success probability.-} Figure\ \ref{distributioncvarvqe} shows the overlap between the exact optimized wave function and the actual solution of QUBO problems with $N = 12$ qubits.  This plot illustrates a qualitatively different behavior in both algorithms. On the one hand, the standard VQE algorithm converges almost always to a classical state, which either coincides with the global minimum or is completely orthogonal to it.  The only exception occurs for the product state \textit{Ansatz} in problems with small connectivity $D \ll 1$.  The CVaR-VQE method, on the other hand, converges to quantum superpositions of classical configurations. This makes the CVaR-VQE wave function have a smaller overlap with the global minimum on average, but it greatly increases the probability of recovering the optimal solution by repeated measurements. In contrast, when the VQE algorithm fails, we have to restart the optimization process again from a different initial condition. The intuition behind this result is that any local minimum of the VQE cost function (\ref{samplemean}) is likely to be a classical state, while any state with an overlap $x$ with the ground state of (\ref{samplemean}) such that $\rho < x$ is a global minimum of the CVaR-VQE cost function (\ref{samplemeanCVaR}). 

\textbf{Success rate and speed: CVaR-VQE is more efficient than standard VQE.-} In Fig.\ \ref{cvarvsvqe} we compare the performance of standard and CVaR-VQE by looking at the success rate, defined by the criterion of Sec. \ref{subsec_efficiencyindicators}, and the number of evaluations.  We investigate QUBO problems with different sizes and densities.  In Figs. \ref{cvarvsvqe}(a) and \ref{cvarvsvqe}(b), we show results with a product state \textit{Ansatz} ($L = 0$), whereas Figs. \ref{cvarvsvqe}(c) and \ref{cvarvsvqe}(d) presents results with $L = 3$ and a linear entanglement pattern.  In both figures we observe that CVaR-VQE clearly outperforms standard VQE, both in terms of success rate and speed, for any choice of problem size, density, and entanglement structure.
Based on these results, from now on the whole study will be carried out considering exclusively the CVaR-VQE algorithm.

\subsection{Entangling layer structure}
\label{subsec_entanglinglayereffect}

Let us now study the influence of correlations in the variational algorithm. Our goal is to understand whether an entanglement structure that imitates the topology of the QUBO problem produces significant advantages. In this section we also introduce the use of a finite number of measurements, exploring how stochasticity affects the different entanglement patterns.

\begin{figure}[b]
\centering
\includegraphics[width=\linewidth]{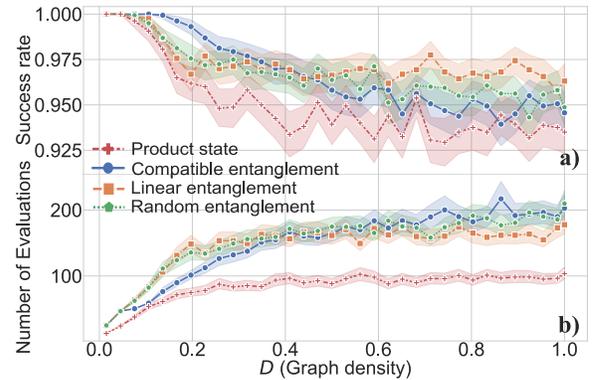}
\caption{Performance of CVaR-VQE $(\rho = 10\%)$ for different entanglement structures and product states, as a function of the QUBO problem's graph density. We simulated problems with $N=12$ qubits, using exact wave functions with $L=0$ or $1$ entanglement layer, and the L-BFGS-B classical optimizer. We plot the average (a) success rate and (b) number of evaluations computed over $N_{\rm ins}=1600$ instances, surrounded by the 95\% confidence interval.}
\label{Entanglinglayer}
\end{figure}

{\bf Shallow entanglement patterns.-}
Figure \ref{Entanglinglayer} illustrates the performance of CVaR-VQE using one or no entanglement layers of the three types mentioned in Sec.\ \ref{sec_ansatzs}. We computed the success rate and function evaluations over 1600 instances of QUBO problems with $N = 12$ qubits, for different values of the associated graph density $D$. We observe that, for small graph densities, a compatible entanglement has a marginal advantage in the success rate and convergence speed. However, as the density of the QUBO graph increases this advantage disappears, and the success rate is matched or outperformed by the classical product state and the linear entanglement wave functions, at slightly lower numbers of function evaluations.

\begin{figure}[b]
\includegraphics[width=\linewidth]{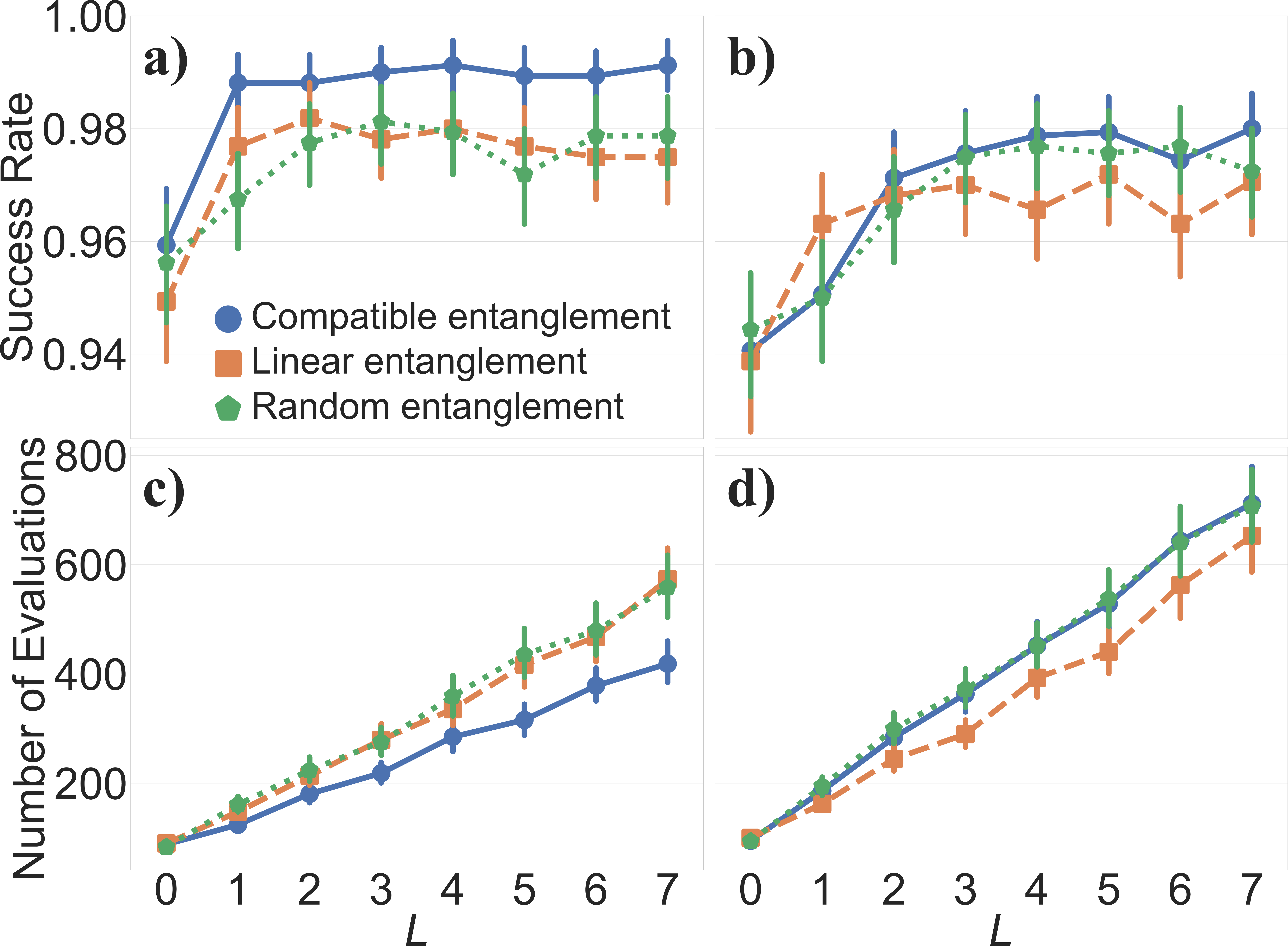}
\caption{Success rate and average number of cost function evaluations needed for convergence, as a function of the number of layers of the variational form. The results show the average of $N_{\rm ins}=1600$ instances and a 95\% confidence interval. Plots (a), (c) and (b), (d) correspond to problems with graph density equal to 0.258 and 0.894, respectively.}
\label{succesvsdepth_en}
\end{figure}

{\bf Efficiency as a function of circuit depth.-}
We have repeated the previous study with increasing number of layers. Figure\ \ref{succesvsdepth_en} illustrates the success rate and number of evaluations, for graphs with intermediate density ($D=0.258$, left column) and dense graphs ($D=0.894$, right column). We observe that in both cases the convergence of the algorithm slows down with the number of layers, proportionally to the growth in the number of parameters. Despite this cost, the success rate reaches a plateau with a few layers, an effect that is especially prominent at smaller densities. In some cases, such as the use of compatible or linear entanglement, the success rate saturates at one layer [cf. Fig.\ \ref{succesvsdepth_en}(a)]. These are shallow enough circuits which have been simulated exactly in a classical computer. Hence, we cannot attribute the effect to vanishing gradients or barren plateaus\ \cite{McClean2018,wang2021noiseinduced}, but to an intrinsic limitation of the algorithm.

\begin{figure}
\includegraphics[width=1\linewidth]{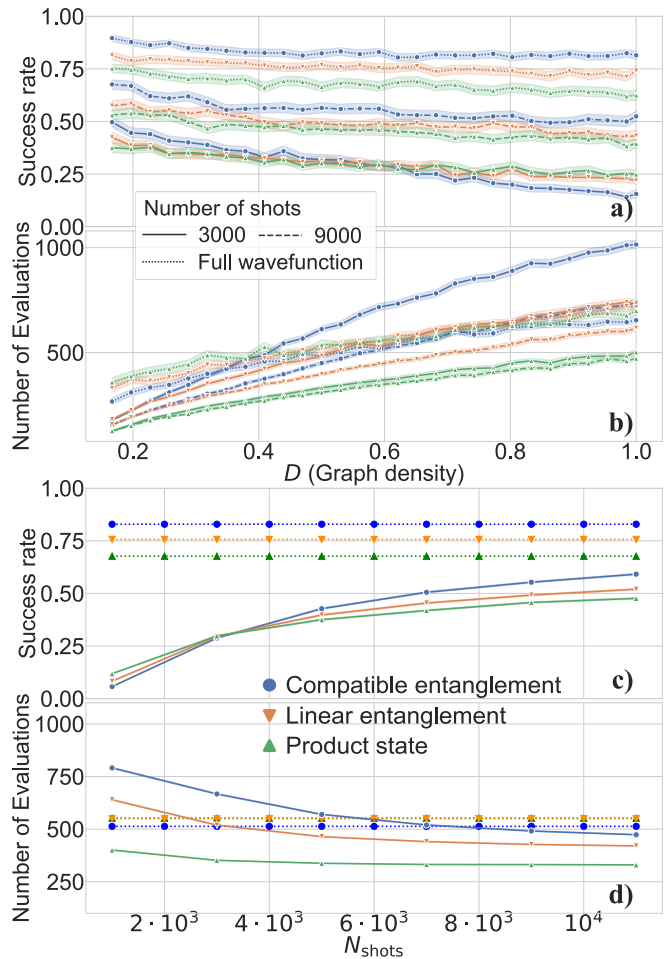}
\caption{Performance of CVaR-VQE $(\rho = 10\%)$ in a realistic simulation with a finite number of measurements, using the SPSA classical optimizer. Calculations have been done using $N=12$ qubits and $L=0,1$ entanglement layers. We plot both the average and a 95\% confidence interval. Plots (a), (b) show the success rate and the speed of convergence respectively as a function of the connectivity of the problem for $3000$ and $9000$ measurements of the variational wave function. Plots (c), (d) illustrate the success rate and speed of convergence as we change the number of measurements, with the ideal wave function simulations $(N_\text{shots}\to\infty)$ in dotted lines.}
\label{Finitesampling}
\end{figure}

{\bf Influence of the number of experimental shots.-}
Until now, our study has focused on exact simulations of the variational wave function, computing the exact expectation values. We will now consider a realistic scenario with a finite number of experimental shots for each observable. In this case, we no longer compute the exact CVaR-VQE cost function, but the random estimator. As explained above, this also requires us to use the SPSA classical optimizer. The results are summarized in Fig.\ \ref{Finitesampling}, for simulations with $L=1$ layers. Note how Figs. \ref{Finitesampling} (a) and \ref{Finitesampling}(b) imitate earlier plots, for different numbers of measurements, while Figs.\ \ref{Finitesampling} (c) and \ref{Finitesampling}(d) focus on a fixed density and study the variation with increasing number of shots.

We can extract several conclusions. First, we see that the compatible entanglement is the most successful \textit{Ansatz} when a sufficient number of shots is available. The contrast with Fig.\ \ref{Entanglinglayer} must be attributed to the change in the optimizer, which also affects the number of evaluations. Second, this advantage is clearly wiped out when we consider a scenario with a finite number of measurements, where the cost function is randomly estimated. Third, we see that the slowdown in convergence produced by the increased graph density is aggravated when the algorithm has incomplete information of the wave function. Third, the previous remark does not apply to the product state \textit{Ansatz}, as discussed in more detail in Sec. \ref{subsec_productstatevsentangledstate}.

{\bf Effect of adapting the entanglement to the problem.-}
Although briefly mentioned in the preceding paragraphs, let us summarize under what circumstances matching the entanglement to the problem structure offers an advantage over other \textit{Ansätze}. With exact wave functions simulations (see Fig. \ref{Entanglinglayer}) and when the information of the cost function is quite incomplete (see 3000 shots case in Fig. \ref{Finitesampling}), compatible entanglement is marginally better than other approaches only when the problem density is low. However, the adaptation of entanglement to the network connections loses its advantageous effect  when the density increases since the specific network structure becomes less noticeable. We also observe a sharper deterioration in the speed of the algorithm when compatible entanglement is applied to high-density problems. If we reduce the uncertainty in the estimation of the objective function raising the number of measurements to 9000 (see Fig. \ref{Finitesampling}), we can see that a graph-adapted entanglement is able to slightly improve the success rate of the algorithm for problems of any density without a significant delay in the convergence.

\subsection{Hardness analysis}
\label{subsec_hardnessanalysis}

At this point we have discussed the efficiency of optimizing QUBO problems according to the characteristics of the variational algorithm such as the cost function or the \textit{Ansatz}. In this section we analyze the problem from a different point of view.  We now try to get some insight into how the specific properties of each QUBO problem affect the hardness of the optimization in terms of success rate and speed of convergence. We define as a hard problem one for which the variational algorithm requires a high number of evaluations to converge, while still obtaining a low success rate.

{\bf Success rate and speed as a function of graph density.-}
We have already concluded from Figs-\ \ref{distributioncvarvqe}-\ref{Entanglinglayer} that the density of the graph severely impacts the performance of the variational algorithm. These studies lead to some general remarks. (a) On average we found that graphs with many connected vertices---i.e., QUBO with dense matrices---are harder problems than sparse problems. (b) When the entanglement structure imitates the graph, we saw that the number of successful instances decreases with the graph connectivity (see Fig. \ref{Entanglinglayer}). (c) However, for variational forms whose design is independent of the problem, we observed that there is a critical density in which the trend steadies around a minimum success rate, despite increasing the number of edges in the graph (see Figs. \ref{cvarvsvqe} and \ref{Entanglinglayer}). Such critical density might be related to the percolation point of the network. (d) We observe equivalent trends when we consider the number of function evaluations.

\begin{figure}
\centering
\includegraphics[width=1\linewidth]{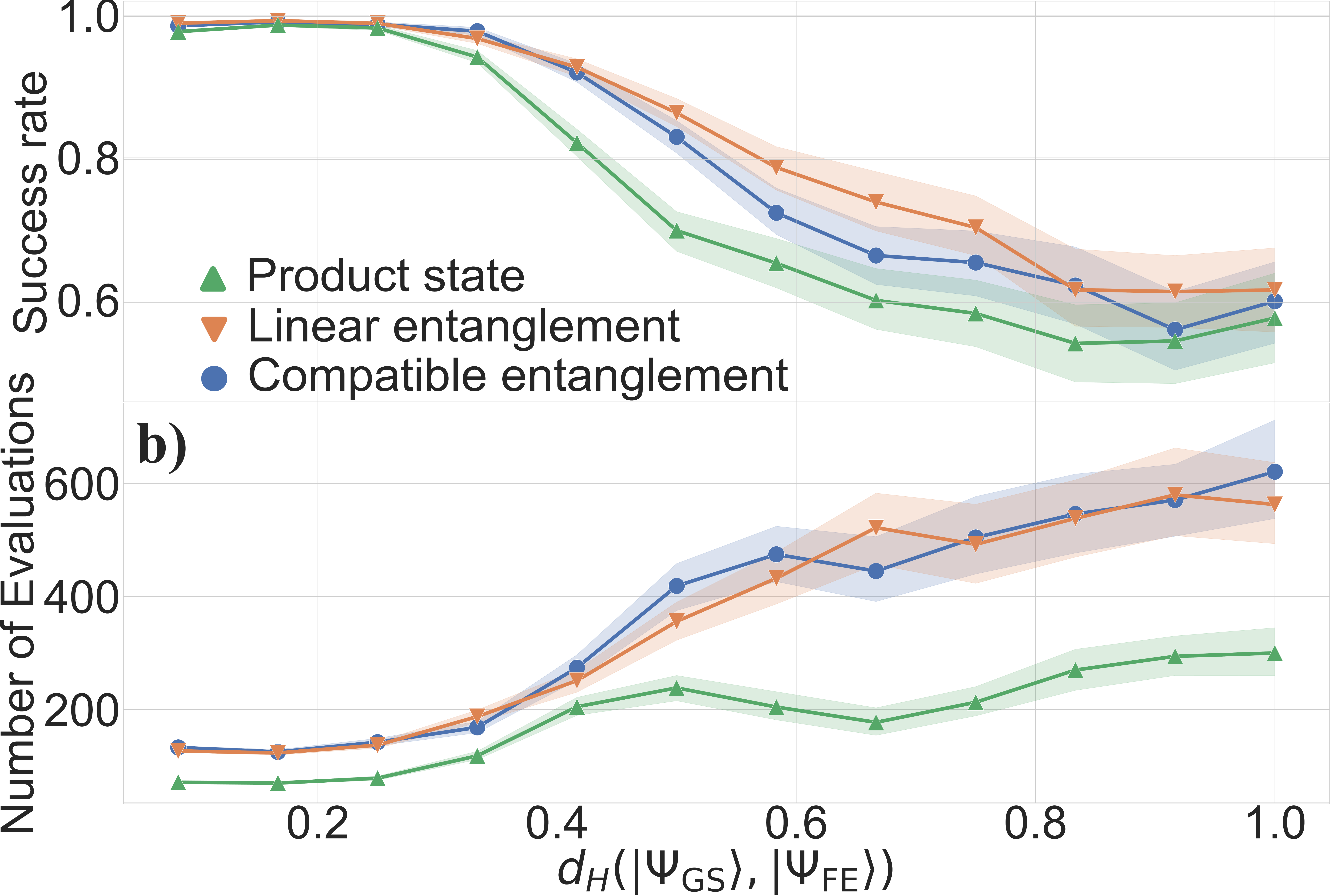}
\caption{Performance of CVaR-VQE $(\rho=10\%)$ as a function of the QUBO's problem Hamming distance between the optimal and the first-excited subspace. We used $N=12$ qubits with $L=0$ or $1$ layers, computing the exact energy with the full wave function $(N_\text{shots}\to\infty).$ The plots show (a) the average success rate and (b) the number of cost function evaluations, with a 95\% confidence interval. }
\label{hardnesssuccessandeval}
\end{figure}

{\bf Impact of low-energy state structure: a measure of hardness.-}
Optimization algorithms typically fail because the \textit{Ansatz} gets trapped in a local minimum that is different from the global minimum we seek. We expect that this phenomenon will be more likely when there are states that have a value of the cost function close to the optimum, but are far from the optimal state in Hilbert space. To test this hypothesis, we have correlated the performance of the variational quantum optimizer to the minimum Hamming distance between the ground state space of the QUBO problem and the (possibly degenerate) manifold of its first excitations. Let us recall that the Hamming distance between two bit strings $\ket{\psi_1} = \ket{s_1, \dots, s_N}$ and $\ket{\psi_2} = \ket{t_1, \dots, t_N}$, is given by $d_{\rm H}(|\psi_1 \rangle, |\psi_2\rangle) = \frac{1}{N} \sum_{j}^{N} |s_j - t_j|.$ We have computed the minimum such distance between the ground state and first-excited manifolds of randomly generated QUBO problem, exploring 29 values of the density and averaging over 1600 random regular graphs for each density.

\begin{figure}
\includegraphics[width=1\linewidth]{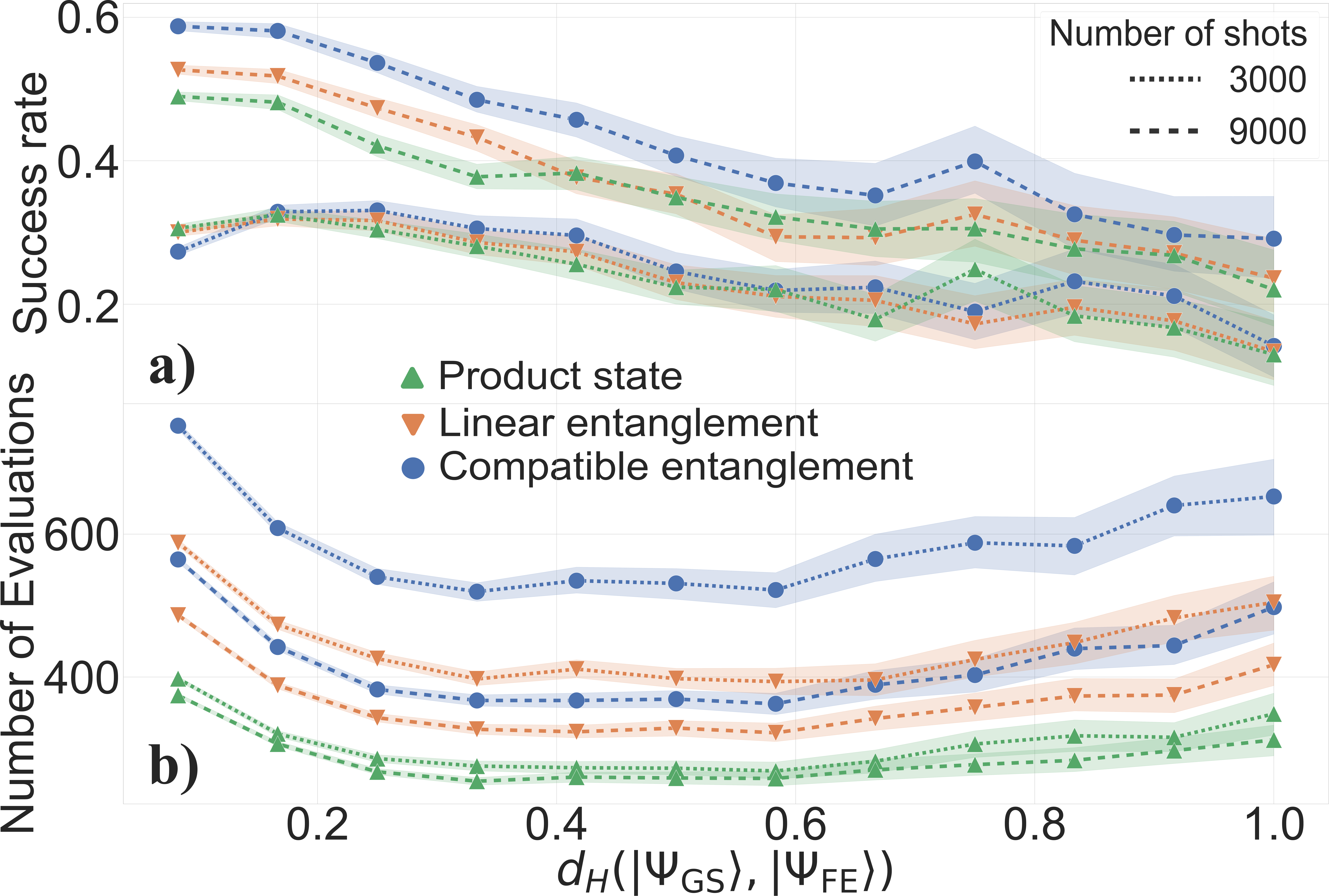}
\caption{Performance of CVaR-VQE $(\rho=10\%)$ as a function of the QUBO problems' Hamming distance between the ground state and lowest-excited manifold, when we consider a limited number of measurements in the estimation of the cost function. Simulations used $N=12$ qubits and $L=0$ and $1$ entanglement layers. We plot the (a) average success rate and (b) average number of evaluations, with a 95\% confidence interval. }
\label{hardnesssuccessandeval_sampling}
\end{figure}

The variational algorithm was simulated using the full wave function and also a finite number of measurements, as shown in Figs. \ref{hardnesssuccessandeval} and \ref{hardnesssuccessandeval_sampling}, respectively. These results show a strong correlation between the Hamming distance from the fundamental state to the first-excited state and the difficulty of the algorithm to solve the optimization problem, both in terms of success rate and speed of convergence. This conclusion holds for all variational \textit{Ans\"atze}.

It must be remarked that the correlation of the difficulty to the Hamming distance is much clearer than to the density (cf. Figs.\ \ref{Entanglinglayer} and \ref{succesvsdepth_en}). While the increase of the graph connectivity led to a maximum loss of success rate of approximately $3\%$, in this case the difference can be of up to $40\%$ for full simulations, and $30\%$ when using 9000-shot measurements. Moreover, the number of objective function evaluations also grows significantly with the Hamming distance. We thus conclude that the Hamming distance is a clearer indicator than the plain graph density to predict the hardness of an optimization problem \cite{Farhi_2008,Callison_2019}.

\subsection{Product state vs. entangled \textit{Ans\"atze}}
\label{subsec_productstatevsentangledstate}

So far we have been considering on equal footing the \textit{Ans\"atze} with $L\geq 1$ entangling layers, and the special case in which we eliminate all entanglers and work with product states. It is important to do a careful comparison between both situations, because only the former case argues for an advantage of using the quantum computer. Indeed, using product states, with CVaR or pure VQE, can be regarded as a new classical optimization algorithm, similar to other methods proposed in the last decade\ \cite{Shin2014, Smolin2014}.

At a glance in Figs. \ref{classicaloptimizers}-\ref{hardnesssuccessandeval_sampling}, we observe a slight improvement in the success rate of VQE when using some entangling operations, as compared with the $L=0$ product state method. This success comes with two costs. First, we need to use a quantum computer, which in itself involves a non-negligible overhead. Second, the increased number of parameters in the variational \textit{Ansatz} also leads to a larger number of function evaluations and slower optimization. Let us analyze these metrics with greater care.

{\bf Product state-based \textit{Ansatz} seems better for few shots.-} The advantage of entangled states decreases and even disappears when we consider the statistical errors induced by a finite number of quantum measurements. This was seen in Figs. \ref{Finitesampling}(a) and \ref{hardnesssuccessandeval_sampling}(a), and more clearly in Fig. \ref{Finitesampling}(b). A decrease in the number of samples reduces the success gap, to a point where at $3000$ shots the product state can outperform other methods. Moreover, as seen in Figs. \ref{Finitesampling}(c) and \ref{hardnesssuccessandeval_sampling}(b), and especially in Fig. \ref{Finitesampling}(d), as we decrease the number of measurements, the product state also saturates in the number of function evaluations, while all other \textit{Ans\"atze} grow. We therefore conclude that the use of a product state variational form (and its fully classical simulation) is advantageous when we are limited in the precision with which we can estimate the cost function.

{\bf The role of entanglement as a function of graph density.-} One may also wonder whether entanglement can provide an advantage in highly- or weakly-connected problems. Inspecting Figs. \ref{Entanglinglayer} and \ref{Finitesampling}, this does not seem to be the case. At large densities with no statistical uncertainty, the success rate is only marginally better in the entangled \textit{Ansatz} case, by $2$-$3\%.$  However, in this region there is a significant performance hit due to the larger number of function evaluations as compared with the product states (see Fig. \ref{Entanglinglayer}). If we consider the stochasticity of quantum measurements, the situation is reversed: entangled \textit{Ans\"atze} achieve greater success rates, by $20\%$ or higher, with faster convergence, regardless of the density of the optimized graph.

{\bf The role of entanglement in different hardness regimes.-}
We have seen that the graph density is not a good quantifier for hardness. It is therefore interesting to look at the correlation between the Hamming distance and the performance of both types of methods, as shown in Table\ \ref{prodvsentan_table}. From inspecting this table and Figs.\ \ref{hardnesssuccessandeval} and \ref{hardnesssuccessandeval_sampling}, we conclude that entanglement provides a moderate advantage in the success rate, ranging from $10\%$ at easy problems, up to $4-5\%$ for hard problems, always in the realistic case with a limited number of measurements. However, the product-state \textit{Ansatz} can be efficiently simulated in the limit of infinite measurements (full wave function). Therefore, in practical terms, we would achieve $25\%$ to $41\%$ higher success rates, with up to an order of magnitude reduction in the number of evaluations, simulating the product-state \textit{Ansatz}.

\begin{table}
\begin{center}
\begin{adjustbox}{max width=\linewidth}
\begin{tabular}{| c | l | c | c | c | c | c | c | c | c | c | }
\cline{3-11}
  \multicolumn{2}{c|}{} & \multicolumn{3}{  c | }{3000 shots} & \multicolumn{3}{ c }{9000 shots} & \multicolumn{3}{ |c| }{Full wave function}\\ \cline{3-11}
  \multicolumn{2}{c|}{~}  & (A) & (B) & (C) &  (A) & (B) & (C) &  (A) & (B) & (C) \\ \hline
\multicolumn{1}{|c|}{\multirow{2}{*}{Success}} & Product &  31\% &  22\% &  16\% & 46\% &  34\% &  26\% &  98\%&  69\% &55\% \\
 & Entangled & 31\% &25\% & 20\% & 57\% & 40\% &30\% & 99\%& 83\% &61\% \\\hline
\multicolumn{1}{|c|}{\multirow{2}{*}{Speed}} & Product & 335 &  274 &  327 &  316  &  262 &  298 &   73 &  206  & 288 \\
& Entangled &496 & 538 & 625 & 463 & 372 &461  & 129 & 390 &560 \\\hline
\end{tabular}
\end{adjustbox}
\caption{Performance of the product-state and entangled-state \textit{Ans\"atze}, for varying hardness and number of shots. The columns (A), (B), and (C) correspond to problems with Hamming distances $d_\text{H}\leq 0.3$, $d_\text{H}\in[0.4,0.7]$, and $d_\text{H}\geq 0.8,$ respectively. The table displays the success rate and the average  number of evaluations of the cost function.}
\label{prodvsentan_table}
\end{center}
\end{table}

Let us sum up the comparison between product-state and entangled \textit{Ansätze}. 
We have found that entanglement yields a slight improvement only in the case of negligible measurement error. 
On the contrary, the product-state \textit{Ansatz} is advantageous, if there is a significant uncertainty in the cost function evaluation (see Figs. \ref{Finitesampling} and \ref{hardnesssuccessandeval_sampling}, and Table \ref{prodvsentan_table}). Our analysis implies that product-state outperforms entangled \textit{Ansätze} in practical terms, since they require fewer resources and can even be classically simulated.

\section{Conclusions}
\label{sec_conclusions}

In this work we have studied the optimization of QUBO problems using the original and CVaR formulations of the VQE algorithm. We have compared the performance of this type of solvers for different variational \textit{Ans\"atze} and for different types of problems, analyzing the success rates, convergence speeds, types of states that are created, and the behavior of the algorithm both in idealized scenarios (unlimited number of measurements) and limited resources.

Our work corroborates the advantage of using a CVaR cost function over conventional averages. We have also verified that adapting the structure entangling operations to the topology of the problem is marginally advantageous, in line with recent experimental results\ \cite{Harrigan2021}. We also found that in a scenario with limited numbers of measurements, entanglement also provides a quantitative advantage over other approaches when enough shots can be performed, but this advantage quickly saturates with the depth of the \textit{Ansatz}. Our study also finds a correlation between the practical hardness of a problem and a metric that characterizes the structure of low-energy excitations.

One important takeaway is that CVaR metrics with product state \textit{Ansatze} provide a classical optimization capable of outperforming the variational quantum eigensolver. This method can afford very large numbers of effective measurements, reaching success rates and performance metrics that exceed those that we see for conventional hybrid quantum-classical methods. This result highlights the need to investigate further the existing algorithms for quantum optimization, including other alternatives such as QAOA\ \cite{farhi2014quantum}. It also illustrates the possibilities that quantum inspired alternatives can bring to the field of classical optimization in the near term.

\subsection*{Acknowledgments}
Work funded by CAM/FEDER project No. S2018/TCS-4342 (QUITEMAD-CM),
Spanish project PGC2018-094792-B-I00 (MCIU/AEI/FEDER, UE), and CSIC Quantum Technology Platform PT-001.
We acknowledge Santander Supercomputacion support group at the University of Cantabria who provided access to the supercomputer Altamira Supercomputer at the Institute of Physics of Cantabria (IFCA-CSIC), member of the Spanish Supercomputing Network, for performing simulations.

\bibliographystyle{apsrev4-2}
\bibliography{bibliography}

\end{document}